\documentclass[10pt]{article}
\usepackage{graphicx}

\begin{document}

\title{THE TWO FORMS OF \\ FRACTIONAL RELAXATION \\ OF DISTRIBUTED ORDER}
\author{
Francesco MAINARDI\thanks{Corresponding author,
E-mail: {\tt francesco.mainardi@unibo.it}},
\ Antonio MURA,
\\ Department of Physics, University of Bologna, and INFN,
\\ Via Irnerio 46, I-40126 Bologna, Italy
\and
Rudolf GORENFLO
\\ Department of  Mathematics and Informatics, Free University of  Berlin,
\\  Arnimallee  3, D-14195 Berlin, Germany
\and
Mirjana STOJANOVI\'C
\\  Department of  Mathematics and Informatics, University of  Novi Sad,
\\  Trg D. Obradovi\'ca 4, 21 000 Novi Sad, Serbia
}

\date{July 2006 - September 2007}

\maketitle

\centerline{International Symposium on Mathematical Methods in Engineering}
\centerline{Ankara, Turkey, April 27-29, 2006}
\centerline{{\bf Journal of Vibration and Control, Vol. 13, pp. 1249-1268  (2007).}}
\begin{abstract}
\noindent
The first-order differential equation of exponential relaxation
can be generalized by using either the fractional.
derivative in the Riemann-Liouville (R-L) sense and
in the Caputo (C) sense,
both of a single order less than 1. The two forms
turn out to be equivalent.
When, however we use fractional derivatives of distributed order
(between zero and 1), the equivalence is lost,
in particular on the asymptotic behaviour of the fundamental
solution at small and large times.
We give an    outline of the theory providing
 the general form of the solution in terms of an integral
 of Laplace type over a positive measure depending on the
 order-distribution.
We  consider with some detail  two cases of fractional
relaxation of distributed order: the double-order   and
the uniformly distributed order discussing
the differences between the R-L and C approaches.
For all the cases considered
we exhibit plots of the solutions
for moderate and large times.
\end{abstract}

\def\pni{\par \noindent}
\def\vsh{\smallskip}
\def\vs{\medskip}
\def\vvs{\bigskip}
\def\vvvs{\bigskip\medskip} 
\def\vsp{\par}
\def\vsn{\vsh\pni}
\def\cen{\centerline}
\def\ra{\item{a)\ }} \def\rb{\item{b)\ }}   \def\rc{\item{c)\ }}
\def\q{\quad} \def\qq{\qquad}
\def \rec#1{{1\over{#1}}}
\def\ds{\displaystyle}
\def\eg{{\it e.g.}\ }
\def\ie{{\it i.e.}\ }
\def\versus{{\it vs.}\ }
\def\e{{\rm e}}
\def\d{\partial}
\def\dx{\partial x}    \def\dt{\partial t}
\def\Ai{{\rm Ai}\,}
\def\Erfc{{\rm Erfc}\,}
\def\u{\widetilde{u}}
\def\ul{\widetilde{u}} 
\def\uf{\widehat{u}} 
\def\r{\right} \def\l{\left}
\def\rt{\right} \def\lt{\left}
\def\lra{\Longleftrightarrow}
\def\RR{\vbox {\hbox to 8.9pt {I\hskip-2.1pt R\hfil}}}
\def\NN{{\rm I\hskip-2pt N}}
\def\CC{{\rm C\hskip-4.8pt \vrule height 6pt width 12000sp\hskip 5pt}}
\def\L{{\mathcal L}} 

\font\bfs=cmbx12 scaled\magstep1

\font\note=cmr10 at 10 truept  
\font\note=cmr8  

\vskip -1truecm
\section{Introduction}
\vskip -0.1truecm
The purpose of this paper is to study two types of fractional
    generalization of the classical relaxation  equation. One type
    uses the fractional derivative in the sense of Riemann and
    Liouville, the other in the sense of   Caputo. In its uses we
    distinguish between single and distributed orders of fractional
    derivatives.

The  plan of the paper is as  follows.
In Section 2, 
 we recall  the relevant properties of the fractional
 relaxation equations of a single order
 $\beta =\beta _1 \in (0,1]$,
 in which  the fractional derivative is intended
 in the Riemann-Liouville (R-L) sense and in the Caputo (C) sense.
 The two forms are shown to be equivalent and the common solutions
corresponding to a few orders are plotted.
In Section 3 we consider two general cases of fractional  
    relaxation where in the R-L setting the order $1 - \beta$,
    in the C-setting the order $\beta$ is distributed according to a
     non-negative weight function $p(\beta)$.
For the fundamental solutions of these equations
 we provide a general formula obtained by the
Titchmarsh theorem on Laplace inversion.
In virtue of this, these solutions appear
as real Laplace transform of a positive spectral
function, and hence they are
 completely monotonic functions for $t \ge 0$
in analogy with the fundamental solution
of  the fractional relaxation equation of a single order.
In Section 4 we consider two  typical  cases of weight function:
the case of two distinct orders $\,0< \beta _1 <\beta _2 \le 1\,$
and the case of uniform distribution of orders between
zero and 1.
For these cases, by using Tauberian theory
 we provide the asymptotic expressions of the fundamental
solution   near zero and near infinity,
that show the different role
played by the order-distribution in the R-L and C approaches.
Finally, concluding remarks 
are given in Section 5.
For the reader's convenience
 we briefly recall in Appendix   the essentials
  of the Fractional Calculus useful for
  understanding the notions of fractional derivative
  in the R-L sense and in the C sense.

\section{Fractional relaxation  of single order}
The classical phenomenon of relaxation  in its simplest form
is known to be governed by  a linear ordinary differential equation
of order one, possibly non-homogeneous,
that hereafter we  recall with the corresponding solution.
Denoting by $t\ge 0 $ the time variable, $u=u(t)$ the field variable,
and  by $\,_tD^1$ the first-order time derivative,
the {\it relaxation} differential equation (of homogeneous type)
reads
$$ \,_tD^1 u(t) = -\lambda  \,u(t)\,, \q t\ge 0\,,  \eqno(2.1)$$
where $\lambda $ is a positive constant denoting the inverse of some
characteristic time. 
The solution of (2.1),
under the initial condition $u(0^+)= 1\,, $ is called
the {\it fundamental solution} and reads
$$ u(t) =  \e^{\ds -\lambda t}\,,
   \q t\ge 0\,.
\eqno(2.2)$$
From the  view-point of Fractional Calculus
(for a short review see the Appendix)  
there appear in the literature two ways of generalizing the
 equation (2.1), one way using the R-L, the other using the
     C fractional derivative.
Adopting the notation of the Appendix
 for the two derivatives, see (A.5)-(A.6), and  denoting
 by $\beta_1$ the common fractional order,
the two forms read respectively for $t\ge 0$
$$ \,_tD^1 \,u(t) = - \lambda \; _tD^{1-\beta} \,u(t)\,, \q 0 < \beta \le 1,
 \eqno(2.3)$$
 and
 $$ \,_tD_*^\beta\, u(t)=  -\lambda \,  u(t)\,, \q 0 < \beta \le 1,
 \eqno(2.4)$$
 where now the positive constant $\lambda$ has dimensions
 $[t]^{-\beta}$.
 If we assume the  same
 initial condition, \eg  $u(0^+)= 1$,
 it is not difficult  to show
 the equivalence of the two forms by
 playing with the operators of standard and fractional integration
 and differentiation\footnote{
 Both Eqs (2.3)-(2.4)
 are equivalent to
 the Volterra integral equation  (of fractional type)
 $$  u(t) = u(0^+) - \lambda  \,_tJ^\beta \, u(t)\,.$$
 For example, we  derive the R-L  equation (2.3) from the fractional
 integral equation
 simply differentiating both sides of the latter,
 whereas we  derive the fractional integral equation
  from the C equation (2.4)  by fractional integration
 of order $\beta$. In fact,
 in view of the semigroup property (A.2) of the 
 fractional integral,   we note that
$$_tJ^\beta\, _tD_*^\beta u(t) \! = \!
  \,_tJ^\beta\, _tJ^{1-\beta}\, _tD^1 \,u(t) \! = \!
  \,_tJ^1\, _tD^1 \,u(t) \!= \! u(t)-u(0^+)\,.$$
In the limit  $\beta =1$ we recover  the relaxation
equation (2.1) with the  solution (2.2).
The reader interested to have more details on  the
two forms of fractional relaxation
may consult, for the R-L approach, the  papers by
Hilfer, Metzler and Nonnenmacher, see \eg
\cite{Hilfer 00a,NonnenMetzler 95},
whereas for the C approach
 the  papers by Caputo, Gorenflo and Mainardi, see \eg
 \cite{CaputoMaina 71,GorMai CISM97,Mainardi CSF96}.
}.

By applying  in Eqs. (2.3)-(2.4) the technique of the Laplace transforms for fractional
derivatives of C and R-L type, see (A.13)-(A.15),
 we get the same result for the fundamental solution, namely
$$ \widetilde {u}(s) =    \frac{s^{\beta-1}}{s^\beta + \lambda}\,,\eqno(2.5)
$$
that, with the  Mittag-Leffler function\footnote{
Let us recall that the Mittag-Leffler function $E_{\beta}(z) $ ($\beta>0$)
is an entire  transcendental function of order $1/\beta\ $, defined
in the complex plane by the power series
$$ E_{\beta} (z) :=
    \sum_{k=0}^{\infty}\,
   {z^{k}\over\Gamma(\beta\,k+1)}\,, \q \beta  >0\,,
 \q z \in\CC\,.  $$
For  details on it we refer  to
\cite{Erdelyi HTF,Kilbas-et-al BOOK06,GorMai CISM97,Podlubny 99,SKM 93}.
We remark that for  $t\ge 0$ the function
$E_\beta (- \lambda t^\beta ) $ preserves the   {\it complete  monotonicity}
of the exponential $\exp(- \lambda t)$: indeed  it
 is represented in terms of a real Laplace transform
of a non-negative  function,
$$ E_\beta (- \lambda t^\beta ) =
   {\displaystyle
   \frac{1} {\pi}\,
   \int _0^\infty \!  \frac{\e^{\,\ds -r t}}{r} \,
   \frac{ \lambda r ^{\beta}    \,\sin (\beta \pi) }
    {\lambda^2 + 2 \lambda \, r ^\beta \,\cos(\beta\pi)+ r^{2\beta} }
        \,dr
        }\,,
  \quad t \ge 0\,,\quad  0<\beta <1\,. $$
However it   decreases at $t \to \infty$ 
like a power  with exponent $-\beta $:
$  E_\beta (-\lambda t^\beta ) \sim {t^{-\beta }}/{[\lambda \Gamma(1-\beta)]}$.
 If $\beta =1/2$
we have for $t\ge 0$:
$ \, E_{1/2} (-\lambda \sqrt{t}) =
    \e^{\ds \,\lambda^2 t}\, \hbox{erfc} (\lambda \sqrt{t})
\sim 1/(\lambda \sqrt{\pi \,t})$
as $t\to \infty$,
where $ \, \hbox{erfc}\,$ denotes the {\it complementary error}
function, see e.g. \cite{AS 65}.}
$\,E_{\beta}$, yields in the time domain
$$ {u}(t) =  E_{\beta} (-\lambda t^\beta)\,,
\q 0<\beta \le 1\,.    \eqno(2.6) $$
We agree to refer to the
equation (2.3) or (2.4)    as the {\it simple fractional relaxation equation}
in the R-L or C sense, respectively.

\newpage
In Fig. 1   we show  
the solution (2.6) 
for a few values of the order $\beta = \beta _1$,
 $\beta_1 = 1/4, 1/2,3/4,1$,
by assuming $\lambda=1$: in the top plate
for the time interval $[0,10]$ (linear scales),
and in the bottom plate
 for the time interval $[10^1, 10^7]$
 (logarithmic scales).
 In the bottom  plate
we have added in  dotted lines the asymptotic values
for $t\to \infty$ in order to better visualize
the power-law decay expressed by $t^{-\beta _1}/\Gamma(1-\beta _1)$
for the cases $0<\beta_1<1$, whereas
the case $\beta_1 =1$ is not visible
in view of the faster exponential decay.
In both plates we have shown
 in dashed line the  singular solution for the limiting case $\beta _1=0$,
 stretching  the definition of the
 Mittag-Leffler function to $E_0(z)= 1/(1-z)$, the geometric series,
 $$   u(t) =
\cases{
E_0(0)= 1 \,, & $t=0\,,$  \cr
E_0(-t^0) \equiv  E_0(-1)= 1/2\,, & $t>0\,,$
} 
\eqno (2.7) $$
\begin{figure}[!]
\begin{center}
\includegraphics[scale=0.30] {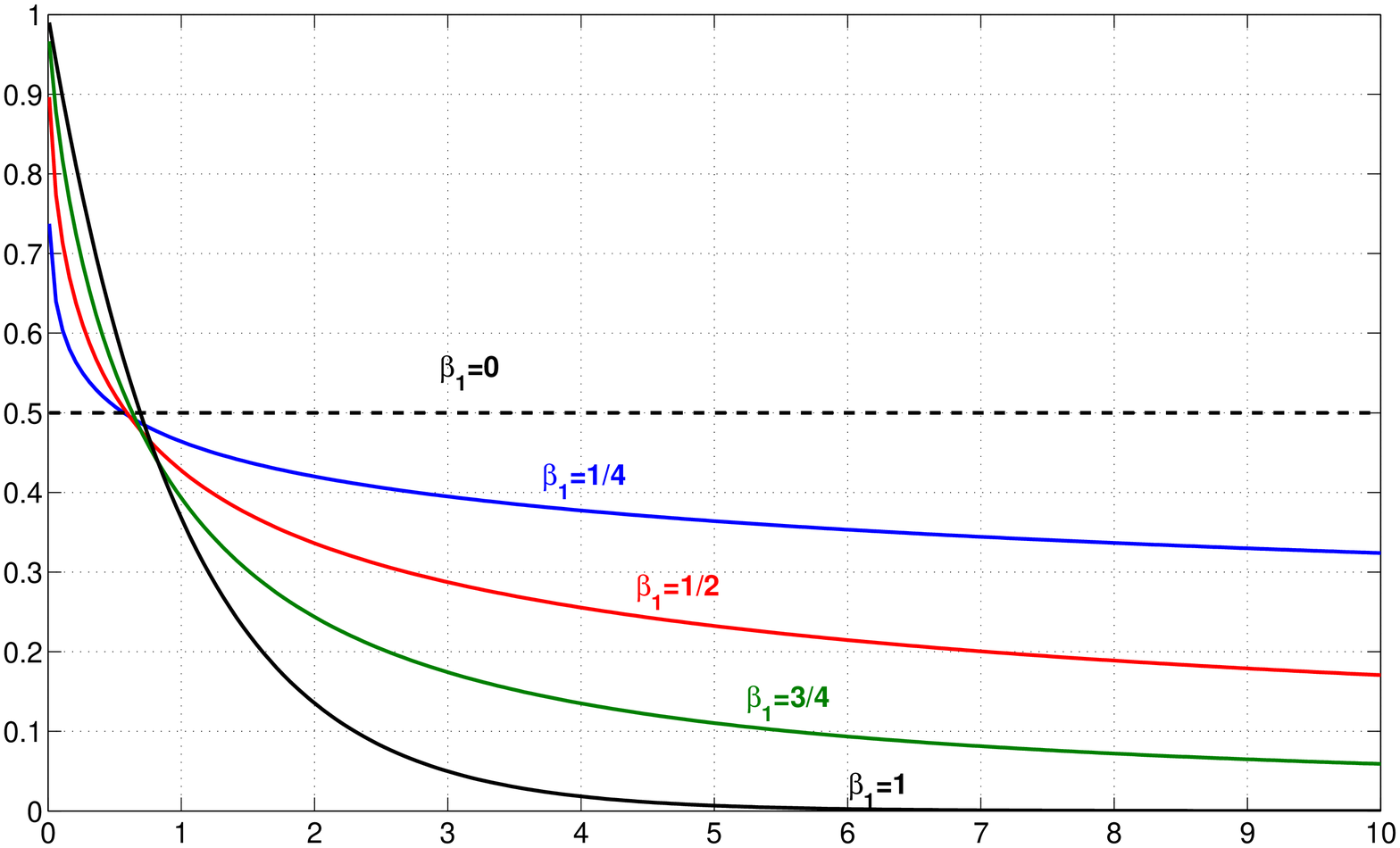}
\includegraphics[scale=0.30] {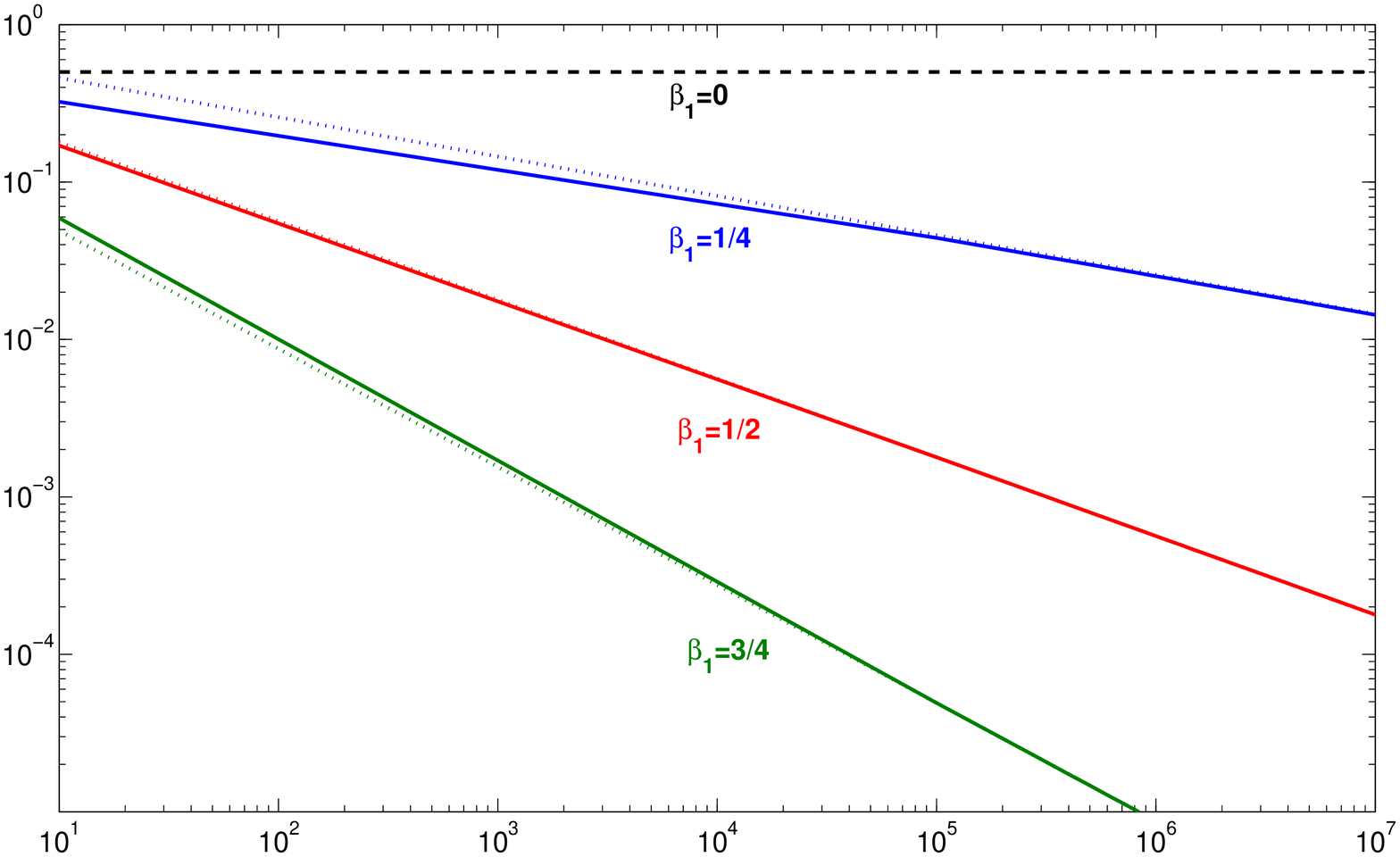}
\end{center}
\vskip -0.5truecm
\caption{Fundamental solutions of the fractional relaxation of a single order
$\beta _1 = 1/4, 1/2, 3/4, 1$. Top: linear scales; Bottom: logarithmic scales.}
\end{figure}

\vfill\eject

\section{Fractional relaxation of distributed order}

\subsection{The two forms for fractional relaxation}

The simple fractional relaxation equations (2.3)-(2.4)
can be generalized by using
the notion  {\it fractional derivative of distributed order}\footnote{
We find a former idea of fractional derivative of distributed order
in time in the 1969 book by Caputo [\cite{Caputo 69},
that was later developed by Caputo himself,
\cite{Caputo FERRARA95,Caputo FCAA01}
and by Bagley \& Torvik,  see \cite{BagleyTorvik 00}.
A basic framework for the numerical solution
of distributed-order differential equations has been recently introduced
by Diethelm \& Ford  \cite{Diethelm-Ford FCAA01},
Diethelm \& Luchko \cite{Diethelm-Luchko 04}
and
by Hartley \& Lorenzo [ \cite{Hartley-Lorenzo SP03,Lorenzo-Hartley ND02}.
}. 
We thus  consider
 the so-called     {\it distributed order
fractional relaxation equation} or
{\it fractional relaxation equation of distributed order},
in the two alternative forms involving the R-L and the C derivatives,
that we write respectively as
$$ \,_tD^1 u(t) \,=\,
-\lambda \, \int _0^1 \! p(\beta )\, _tD^{1-\beta} u(t) \, d\beta\,,
\eqno(3.1)
$$
and
$$
  \int _0^1 \! p(\beta )\, _tD_*^\beta u_*(t) \, d\beta
\,    =  -\lambda \,u_*(t) \,,
\eqno(3.2) $$
subjected to the  initial condition $u(0^+) =u_*(0^+) = 1$,
where
$$  p(\beta)\ge 0\,, \q \hbox{and} \q
   \int _0^1 \! p(\beta )\,d\beta =  c >0\,. \eqno(3.3)$$
The positive constant $c$ can be taken as 1
if we want the integral to be   normalized.
For the weight function
   $p(\beta)$
   we    conveniently  require  
that its primitive
$ P(\beta) = \int_0^\beta \!\!   p(\beta^\prime)\, d \beta^\prime $
vanishes at $\beta=0$ and is there continuous from the right,
attains the value $c$ at $\beta=1$
and has at most finitely many  (upwards) jump points in 
the half-open interval $0<\beta\le 1$,
these jump points allowing delta contributions to 
$p(\beta)$  
(particularly relevant for discrete distributions of orders).

Since for distributed order the solution depends
on the selected approach (as we shall show hereafter),
we now  distinguish the fractional equations (3.1) and (3.2)
and their fundamental solutions
by decorating in the Caputo case
the variable $u(t)$ with subscript $*$.

As in  \cite{GorMai FRACTAL06}
the present analysis  is  based on the application of the Laplace transformation
with particular attention to some  special cases. Here, for these cases, we shall
provide plots of the corresponding solutions.

\subsection{The  integral formula for the fundamental solutions}

Let us now apply the Laplace transform to Eqs. (3.1)-(3.2) by using
the  rules (A.15) and (A.13) appropriate to the R-L and C derivatives,
respectively.
Introducing the relevant functions
$$ A(s) = s \int _0^1 \! p(\beta )\, s^{-\beta} \,d\beta  \,, \eqno(3.4)$$
and
$$ B(s) =  \int _0^1 \! p(\beta )\, s^\beta \,d\beta  \,, \eqno(3.5)$$
we then get for the R-L and C cases, after simple manipulation,
the Laplace transforms of the corresponding fundamental solutions:
$$  \widetilde u(s) =
\frac{ 1}{s + \lambda  A(s)}\,,
\eqno(3.6)$$
and
$$  \widetilde u_*(s) =
\frac{ B(s)/s}{\lambda + B(s)}\,.
\eqno(3.7)$$
We easily note that in the particular case $p(\beta)= \delta(\beta -\beta _1)$
we have in (3.4): $A(s)= s^{1-\beta _1}$, and in (3.5): $B(s)= s^{\beta _1}$.
Then,  Eqs. (3.6) and (3.7) provide the same result (2.5) of the
simple fractional relaxation.

By inverting  the  Laplace transforms in (3.6) and (3.7)
 we obtain the  fundamental solutions 
 for the R-L and C fractional relaxation of distributed order.

Let us start with the R-L derivatives.
We get (in virtue of  the Titchmarsh theorem on Laplace inversion)
the representation
$$   {u} (t) = -\rec{\pi }\,
\int_0^\infty \! \e^{-rt} \,  \hbox{Im}
\lt\{\widetilde {u} \lt(r \e^{i\pi }\rt)\rt\}
\, dr\,, \eqno(3.8) $$
that requires the expression of
$ - \hbox{Im} \lt\{1/[s + \lambda A (s)]\rt\}$
along the ray $s=r\,\e^{i\pi }$ with $r>0$
(the branch cut of the function   $s^{-\beta} $).
We write 
$$ A \lt(r\, \e^{\,\ds i \pi}\rt) =
\rho \,\cos (\pi \gamma)
+ i \rho \sin (\pi \gamma)\,, \eqno(3.9)$$
where
$$
\cases{
{\ds  \rho =\rho (r) =\left\vert A\lt(r \,\e^{i\pi }\rt) \right\vert}\,, \cr
{\ds \gamma = \gamma (r) =
\rec{\pi}\,\hbox{arg}  \left[A\lt (r \,\e^{i\pi }\rt)\right]}\,.
} \eqno(3.10)$$
Then,  after simple calculations,  we get
$$  u(t) =
\int_0^{\infty}\! \e^{\, \ds -rt}\,H  (r;\lambda) \,dr \,, \eqno(3.11)$$
with
$$ H(r;\lambda )
=    \frac{1}{\pi }\,
\frac{ \lambda \,  \rho \,\sin (\pi \gamma)}
{r^2 - 2 \lambda \, r\, \rho\,\cos (\pi \gamma) +\lambda^2 \rho^2}
\ge 0 \,. \eqno(3.12)$$
 
 Similarly for the C derivatives
 we obtain
$$   {u}_* (t) = -\rec{\pi }\,
\int_0^\infty \! \e^{-rt} \,  \hbox{Im}
\lt\{\widetilde {u}_* \lt(r \e^{i\pi }\rt)\rt\}
\, dr\,, \eqno(3.13) $$
that requires the expression of
$ - \hbox{Im} \lt\{ B(s)/[s (\lambda +B (s))]\rt\}$
along the ray $s=r\,\e^{i\pi }$ with $r>0$
(the branch cut of the function   $s^{\beta} $).
We write
$$ B \lt(r\, \e^{\,\ds i \pi}\rt) =
\rho _* \,\cos (\pi \gamma _*)
+ i \rho _* \sin (\pi \gamma _*)\,, \eqno(3.14)$$
where
$$
\cases{
{\ds  \rho_* =\rho_* (r) =\left\vert B\lt(r \,\e^{i\pi }\rt) \right\vert}\,, \cr
{\ds \gamma_* = \gamma_* (r) =
\rec{\pi}\,\hbox{arg}  \left[B\lt (r \,\e^{i\pi }\rt)\right]}\,.
} \eqno(3.15)$$
 After simple calculations  we get
$$  {u_*}(t) =
\int_0^{\infty}\! \e^{\, \ds -rt}\,K  (r;\lambda) \,dr \,, \eqno(3.16)$$
with
$$ K(r;\lambda )
=    \frac{1}{\pi \,r}\,
\frac{ \lambda\rho_* \,\sin (\pi \gamma_*)}
{\lambda^2 + 2 \lambda \, \rho_*\,\cos (\pi \gamma_*) +\rho_*^2} \ge 0
\,. \eqno(3.17)$$
We note from (3.11) and (3.16)
that, since $H(r; \lambda)$ and $K(r;\lambda)$  are  non-negative  functions
 of $r$ for any $\lambda \in \RR^+$,
the fundamental solutions  $u(t)$ and $u_*(t)$
keep the relevant property  to be { \it completely monotone}.

The integral expressions (3.11) and (3.16) 
provide  a sort of spectral representation of the fundamental solutions
that  will be  used to  numerically evaluate
 these solutions
in some examples considered as interesting cases.

Furthermore, it is quite instructive to    compute
for the fundamental  solutions
their asymptotic expressions for $t\to 0$ and $t\to \infty$
because they provide
their  analytical (even if approximated) representations
for sufficiently short and long time respectively,
and useful checks for the numerical evaluation
in the above time ranges.

To derive these asymptotic representations we shall
apply the Tauberian theory of Laplace transforms.
According to this theory
 the asymptotic behaviour of a function $f(t)$ near $t=\infty$
and $t =0$
is (formally)  obtained from the asymptotic behaviour
of its Laplace transform $\widetilde f(s)$ for $s \to 0^+$ and
for $s \to +\infty$, respectively.
For this purpose  we note the asymptotic representations,
\\ from (3.6):
$$\widetilde  u(s)   
\sim \cases{
{\ds \frac{1}{\lambda \,A(s)}}\,,&
$s \to 0^+ \,,\quad
 \hbox{being}\quad  A(s)/s >> \lambda \,,$
\cr
{\ds \rec{s}\lt[1- \lambda \, \frac{A(s)}{s}\rt]}\,, &
$s \to +\infty \,,\quad \hbox{being}\quad A(s)/s << 1/\lambda \,,$
}
\eqno(3.18)
$$
and from (3.7):
$$\widetilde  u_*(s)   
\sim \cases{
{\ds \rec{\lambda }\,   \frac{B(s)}{s}}\,,&
$s \to 0^+ \,,\quad
 \hbox{being}\quad B(s) << \lambda \,,$
\cr
{\ds \rec{s}\lt[1- \frac{\lambda }{B(s)}\rt]}\,, &
$s \to +\infty\,, \quad
  \hbox{being}\quad B(s) >> \lambda \,.$
}
\eqno(3.19)
$$

\newpage

\section{Examples}
Since   finding explicit solution formulas
is not possible for the relaxation equations 
(3.1) and (3.2)
we shall concentrate our interest
to some typical choices for the  weight function $p(\beta )$
in (3.3) that characterizes the order distribution.
For these choices we present
the numerical evaluation
of the Titchmarsh integral formula, see Eqs (3.8)-(3.12)
for $u(t)$ (the R-L case), and Eqs. (3.13)-(3.17)
for $u_*(t)$ (the C case).
The numerical results  are  checked
by  verifying the matching
with  the asymptotic expressions
of $u(t)$ and $u_*(t)$  as $t\to 0$  and $t\to +\infty$,
obtained via
the Tauberian theory for  Laplace transforms., according
to Eqs. (3.18)-(3.19).

\subsection{The double-order fractional relaxation}

We now consider the choice
$$
p(\beta ) =p_1\delta (\beta -\beta_1 )+ p_2\delta (\beta -\beta_2)\,,
 \quad
 0<\beta _1<\beta _2\le 1\,,
 \eqno(4.1)$$
where the constants $p_1$ and $p_2$ are both positive,
conveniently  restricted to the normalization condition $p_1+p_2=1$.
Then for the R-L case we have
$$
A(s) = p_1 \,s^{1-\beta _1} + p_2\, s^{1-\beta _2}\,, \eqno(4.2)
$$
so that, inserting (4.2) in  (3.6),
$$  \widetilde u(s) =
\frac{ 1}{s[1 + \lambda (p_1 \,s^{-\beta _1} + p_2\, s^{-\beta _2})]}
\,, \eqno(4.3)
$$
Similarly, for the C case we have
$$
B(s) = p_1 \, s^{\beta _1} + p_2 \,s^{\beta _2} \,,
\eqno(4.3)
$$
so that, inserting (4.3) in (3.7),
$$  \widetilde u_*(s) =
\frac{ p_1 \, s^{\beta _1} + p_2 \,s^{\beta _2}}
{s[\lambda + p_1 \,s^{\beta _1} + p_2\, s^{\beta _2}]}
\,. \eqno(4.4)$$
We leave as an exercise the derivation of the spectral functions
$H(r;\lambda)$ and $K(r;\lambda)$
of the corresponding fundamental solutions, that are used
for the numerical  computation. The numerical results
are checked by their  matching  with the
asymptotic expressions that we  evaluate
by invoking the Tauberian theory and using Eqs. (3.18)-(3.19)
jointly  with Eqs (4.2)-(4.3) respectively.

 For  the R-L-case
 we note  that in (4.2)  ${\ds s^{1-\beta_1}}$ is  negligibly
small in comparison with  ${\ds s^{1-\beta_2}}$for $s \to 0^+$ and, viceversa,
   ${\ds s^{1-\beta_2}}$ is  negligibly small
in comparison to ${\ds s^{1-\beta_1}}$for $s \to +\infty$.
Similarly for the C-case
we note  that in (4.3) ${\ds s^{\beta_2}}$ is negligibly small
in comparison to ${\ds s^{\beta_1}}$for $s \to 0^+$ and, viceversa,
   ${\ds s^{\beta_1}}$ is negligibly small
in comparison  ${\ds s^{\beta_2}}$for $s \to +\infty$.

\noindent
As a consequence of these  considerations
we get  for the R-L case, if $\beta_2<1$,
$$
\widetilde u (s) \,\sim\,
 \cases{
 {\ds \frac{1}{\lambda\, p_2 }\, s^{\beta _2-1}}\,, &
                         $s \to 0^+ \,,$ \cr
  {\ds \rec s \lt  (1 - \lambda \, p_1\, s^{-\beta _1} \rt)} \,, &
                        $s \to + \infty\,,$
}
\eqno(4.5)$$
so that
$$ u(t) \,\sim \,
\cases{
{\ds \frac{1}{\lambda\, p_2}\,\frac{t^{-\beta _2}}{\Gamma(1-\beta _2)} }\,,&
 $t \to +\infty\,,$ \cr
{\ds 1- \lambda \, p_1\, \frac{ t^{\beta _1}}{\Gamma(1+\beta _1)}}\,,&
 $t \to 0^+\,.$
}
\eqno(4.6)$$
\begin{figure}[!h]
\begin{center}
\includegraphics[scale=0.30]{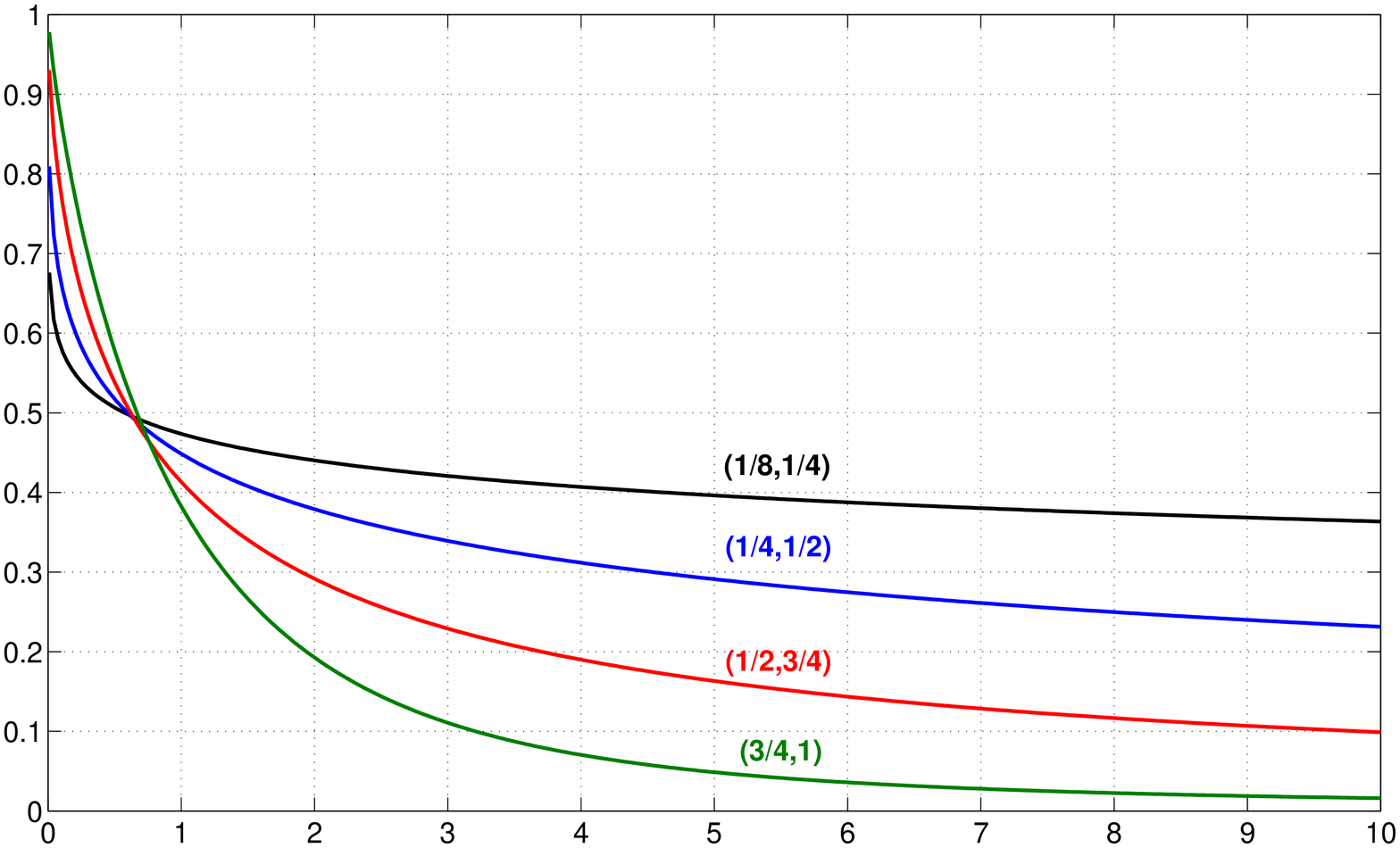}
\includegraphics[scale=0.30]{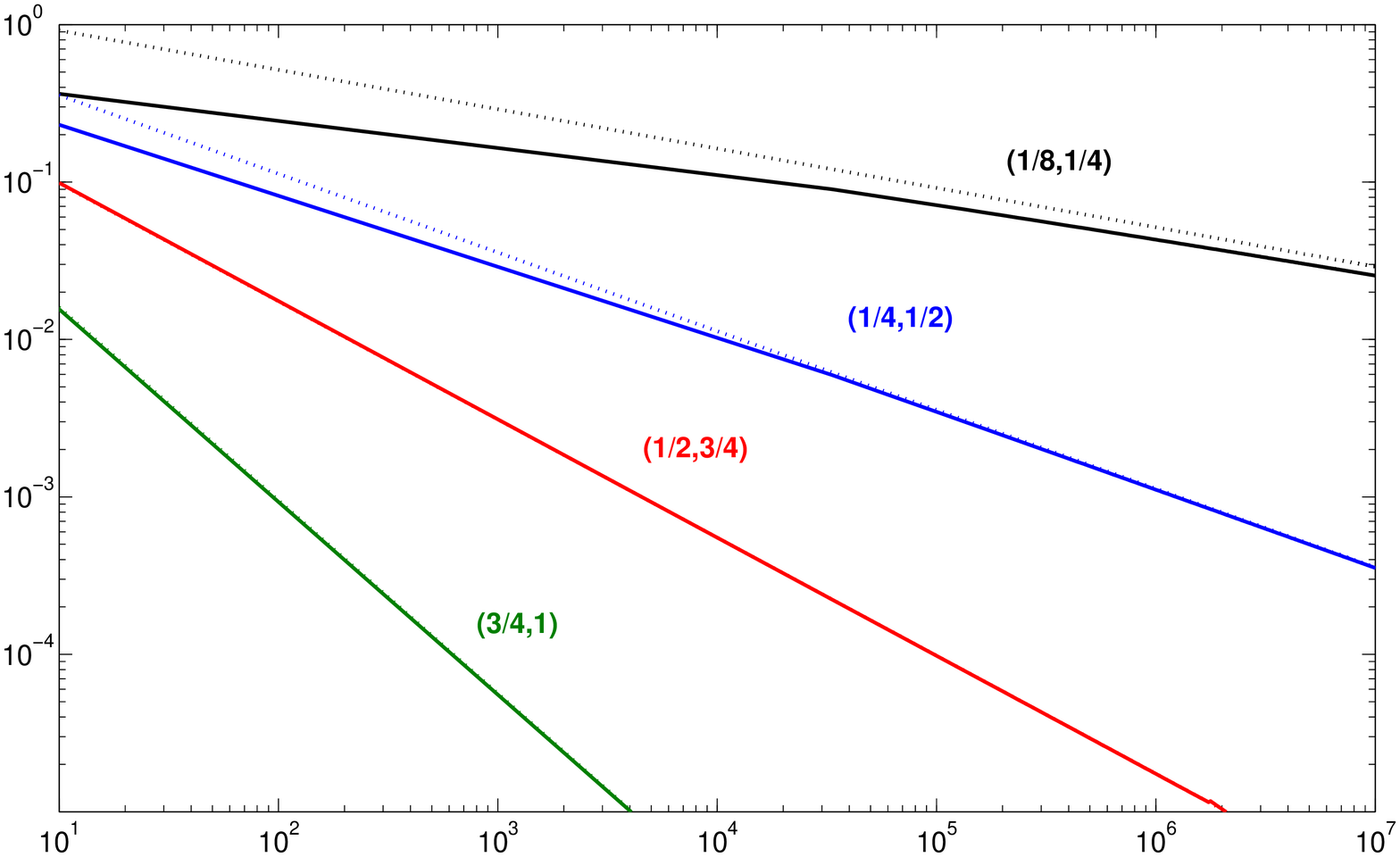}
 \vskip -0.5truecm
\caption{Fundamental solutions of the R-L-fractional relaxation of double order
in some $\{\beta _1,\beta _2\}$ combinations:
$\{1/8, 1/4\}; \, \{1/4,1/2\};\, \{1/2,3/4\}; \,\{3/4,1\}$.
Top: linear scales; Bottom: logarithmic scales.}
\end{center}
\end{figure}

We note that 
 Eq. (4.5a) and henceforth Eq. (4.6a)
lose their meaning for $\beta _2=1$.
 In this case we need 
 a more careful reasoning:
we consider
the  expression  for $s\to 0$  provided by
(3.18) as it stands, that is
 $$ \widetilde u(s) \sim \frac{1}{\lambda\,[p_1 s^{1-\beta _1} + p_2]}
 =    \frac{1}{\lambda \,p_1}\, \frac {1}{s^{1-\beta _1}+ p_2/(\lambda \,p_1) }
 \,. \eqno (4.7)$$
In virtue of the Laplace transform pair
 $$      t^{\nu -1} E_{\mu ,\nu }\lt(- q t^\mu\rt) \div
 \frac{s^{\mu -\nu} }{s^\mu + q}\,,\eqno(4.8)$$
 see Eq. (1.80) in \cite{Podlubny 99},
 where $E_{\mu, \nu}$ denotes the Mittag-Leffler function
 in two parameters\footnote{
 The Mittag-Leffler function 
$E_{\mu,\nu}(z) $ ($ \Re \{\mu\} >0$,  $\, \nu \in \CC$)
is  defined 
by the power series
$$ E_{\mu, \nu} (z) :=
    \sum_{k=0}^{\infty}\,
   \frac{z^k}{\Gamma(\mu \,k + \nu)}\,,
 \q z \in\CC\,.  $$
 It generalizes the classical Mittag-Leffler function
 to which it reduces for $\nu=1$.
 It is an entire  transcendental function of order $1/\Re \{\mu\} $
  on which the reader can inform himself  by again consulting  
\cite{Erdelyi HTF,Kilbas-et-al BOOK06,GorMai CISM97,Podlubny 99,SKM 93}.
 With  $\mu, \nu \in \RR $ the function $E_{\mu, \nu}(-x)$ ($x\ge 0$)
 turns a completely monotonic function of $x$ if $0< \mu \le 1$
 and $\nu \ge \mu$, see \eg \cite{Miller-Samko 01}. This property is still valid
 when $x = q \,t^\mu$ ($q>0$).
 In particular, for $0<\mu=\nu<1$ we note
 $$ q \,t^{-(1-\mu )}
  \, E_{\mu ,\mu} \lt(- q\, t^{\mu }\rt)
=  -\frac{d}{dt} E_\mu  \lt (-q\, t^{\mu }\rt)
\sim  \frac {\mu}{q\, \Gamma(1-\mu)}\, t^{-(\mu+1)}\,,
\quad t \to +\infty\,.$$
 }
we get, with $q= p_2/(\lambda \,p_1)$ and $\mu = \nu = 1-\beta _1$,
as $t\to +\infty\,:$
 $$u (t) \sim
{\ds \frac{1}{\lambda \,p_1}}\,
{\ds t^{-\beta _1}}\,
{\ds  E_{1-\beta _1,1- \beta _1 } \lt(- q t^{1 -\beta _1}\rt)}
= - {\ds \frac{1}{\lambda \,p_1}}\,
{\ds \frac{d}{dt} E_{1-\beta _1} \lt(-q t^{1-\beta _1}\rt)}\,.
  \eqno(4.9)
 $$
Taking into account  the asymptotic behaviour of  the Mittag-Leffler
function, we finally get
$$ u(t) \sim
\lambda \, \frac{p_1}{p_2}\, \frac{1- \beta _1}{\Gamma(\beta_ 1)}\,t^{\ds \,-(2-\beta _1)}
\q \hbox{as} \q t\to +\infty\,.\eqno (4.10)$$

Similarly for the C case we get:
$$
\widetilde u_* (s) \,\sim\,
 \cases{
 {\ds \frac{p_1}{\lambda }\, s^{\beta _1-1}}\,, &
                         $s \to 0^+ \,,$ \cr
  {\ds \rec s \lt  (1 - \frac{\lambda}{p_2} s^{-\beta _2} \rt)} \,, &
                        $s \to + \infty\,,$
}
\eqno(4.11)$$
so that
$$
u_*(t) \,\sim \,
\cases{
{\ds \frac{p_1}{\lambda}\,\frac{t^{-\beta _1}}{\Gamma(1-\beta _1)}}\,,&
 $t \to +\infty\,,$ \cr
{\ds 1- \frac{\lambda}{p_2}\, \frac {t^{\beta _2}}{\Gamma(1+\beta _2)}}\,,&
 $t \to 0^+\,.$
}
\eqno(4.12)$$
\begin{figure}[!h]
\begin{center}
\includegraphics[scale=0.30]{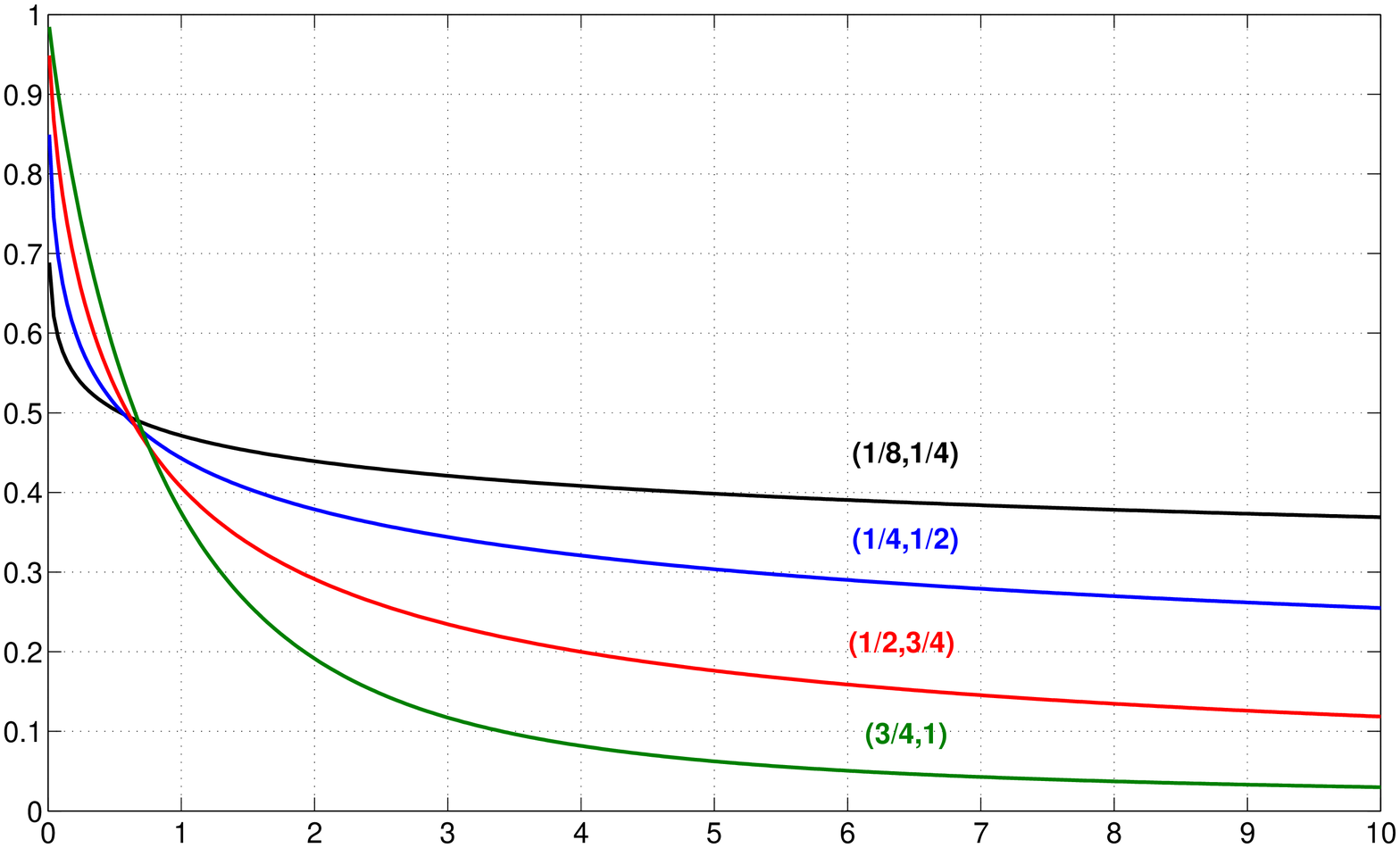}
\includegraphics[scale=0.30]{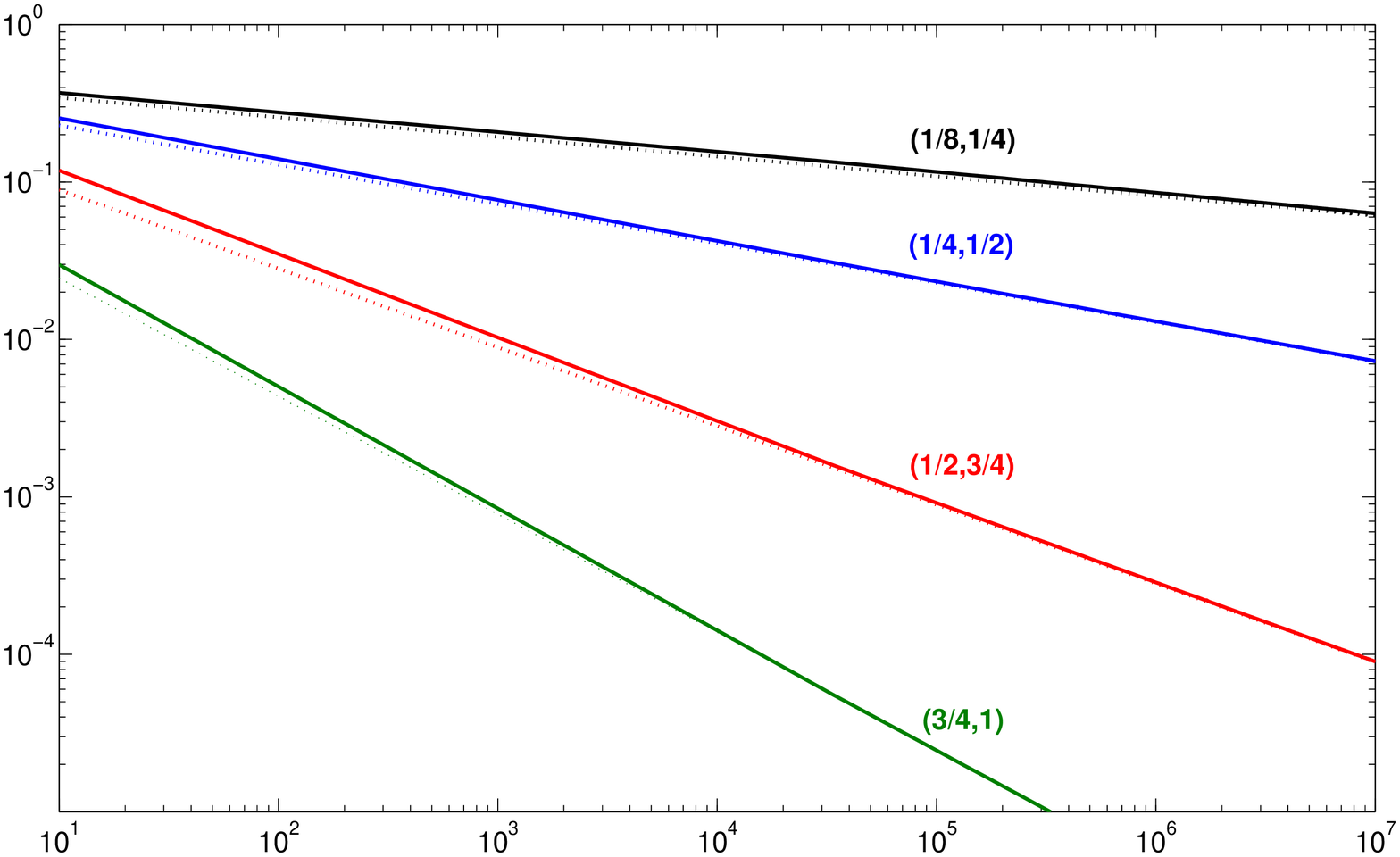}
 \vskip -0.5truecm
\caption{Fundamental solutions of the C-fractional relaxation of double order
in some $\{\beta _1,\beta _2\}$ combinations:
$\{1/8, 1/4\}; \, \{1/4,1/2\};\, \{1/2,3/4\}; \,\{3/4,1\}$.
Top: linear scales; Bottom: logarithmic scales.}
\end{center}
\end{figure}
We exhibit in Figs. 2-3 the
plots of the fundamental solutions for
R-L and C fractional relaxation, respectively,
in some $\{\beta _1,\beta _2\}$ combinations:
$\{1/8, 1/4\}$; $\, \{1/4,1/2\}$; $\, \{1/2,3/4\}$;
$\,\{3/4,1\}$.
We  have  chosen $p_1=p_2=1/2$ and, as usual $\lambda=1$.
From the plots  the reader is expected to verify the role
played by the different orders for small and large times
according to the corresponding asymptotic expressions,
see Eqs. (4.6), (4.9)-(4.10) and (4.12). 

\subsection{The uniformly distributed order fractional relaxation}

We now consider the choice
$$
p(\beta ) =1\,, \q 0 <\beta < 1\,.  \eqno(4.13)$$
For the R-L case we have
$$A(s) = s \! \int_0^1 \! s^{-\beta} \, d\beta  = \frac{s-1}{\log s}\,, \eqno(4.14)$$
hence, inserting (4.14) in (3.6)
$$  \widetilde u(s)=
\frac{\log s}{s\log s  + \lambda \,(s-1)} \,.
\eqno (4.15)
$$
For the C case    we have
$$B(s) = \int_0^1 \! s^\beta \, d\beta  = \frac{s-1}{\log s}\,, \eqno(4.16)$$
hence, inserting (4.16) in (3.7),
$$
\widetilde u_*(s)
= {\ds \rec{s} \frac{s-1}{\lambda \log s +s -1}}
=
   {\ds  \rec{s} - \rec{s}\frac{\lambda  \, \log  s}{\lambda  \, \log  s+s-1}}\,.
\eqno(4.17)$$
We note that for  this special order distribution
we have  $A(s) = B(s)$
but the corresponding fundamental solutions are quite different,
as we see from their Laplace transforms (4.15) and (4.17).

Then, invoking the Tauberian theory for   {\it regularly
varying    functions}  (power functions multiplied by
{\it slowly varying functions}\footnote{
{\bf Definition:} We call a (measurable) positive function $a(y)$,
defined in a right neighbourhood of zero, {\it slowly varying at zero} if
$a(cy)/a(y) \to 1$ with $y \to 0$ for every $c>0$.
We call a (measurable) positive function $b(y)$,
defined in a  neighbourhood of infinity, {\it slowly varying at infinity}
if
$b(cy)/b(y) \to 1$ with $y \to \infty$ for every $c>0$.
Example: $|\log y|^{\gamma}$ with $\gamma \in \RR$.}, 
 a topic
adequately treated  in the treatise on Probability
 by Feller  \cite{Feller 71},
 Chapter XIII.5, we have the following asymptotic
 expressions for the R-L and C cases.

 For the R-L case we get
$$
\widetilde u (s) \,\sim\,
 \cases{
 {\ds \frac{\log s}{\lambda (s-1)}}\,, &
$s \to 0^+ \,,$
\cr
{\ds \rec{s} \lt[1- \lambda\,\frac{s-1}{s \log s}\rt]} \,, &
 $s \to + \infty\,,$
}
\eqno(4.18)$$
so
$$
u(t) \,\sim \,
\cases{
{\ds \rec{\lambda}\, \e^{\ds \, t}\,\mathcal{E}_1(t) \sim
\frac{1}{\lambda\, t}} \,,&
 $t \to +\infty\,,$ \cr\cr
 {\ds 1 - \frac{\lambda} {\lt\vert \log (1/t)\rt \vert}}\,,&
 $t \to 0^+\,.$
}
\eqno(4.19)$$
In (4.19a) 
$\mathcal{E}_1(t) :=
{\ds \int_t^\infty \! \frac{\e^{\ds -u}}{u}\, du}$
denotes the exponential integral,
 see \cite{AS 65}, Ch. 5 and the Laplace transform pair (29.3.100). 

 For the C case we get
$$
\widetilde u_* (s) \,\sim\,
 \cases{
 {\ds \frac{1}{\lambda\, s \log (1/s) }}\,, &
$s \to 0^+ \,,$
\cr
{\ds \rec{s} - \frac{\lambda\, \log s}{s^2}} \,, &
 $s \to + \infty\,,$
}
\eqno(4.20)$$
so
$$
u_*(t) \,\sim \,
\cases{
{\ds \frac{1}{\lambda\,\log t}}\,,&
 $t \to +\infty\,,$ \cr
{\ds 1- \lambda \,t \log (1/t)}\,,&
 $t \to 0^+\,.$
}
\eqno(4.21)$$
\vfill\eject
\begin{figure}[!h]
\begin{center}
\includegraphics[scale=0.30]{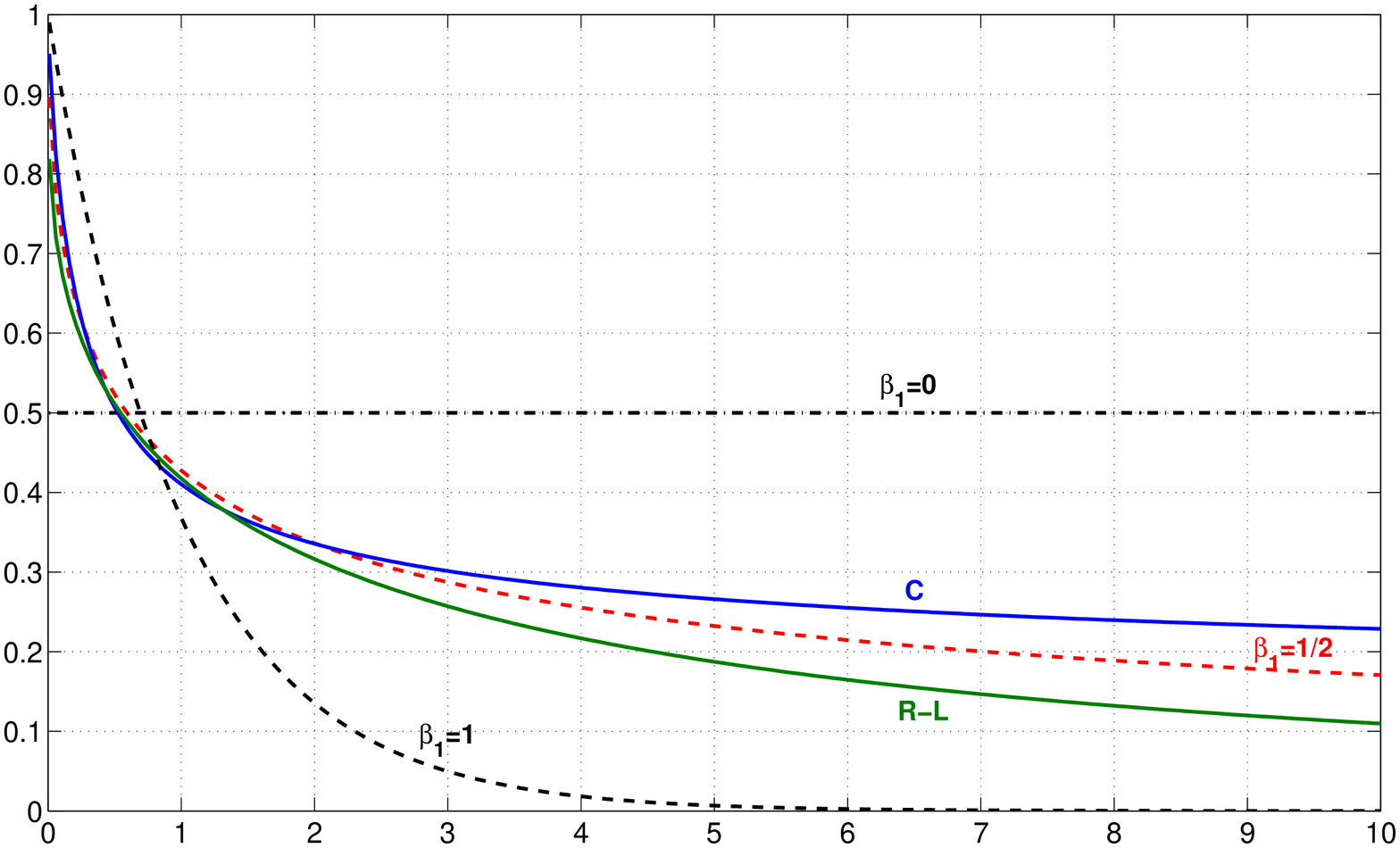}
\includegraphics[scale=0.30]{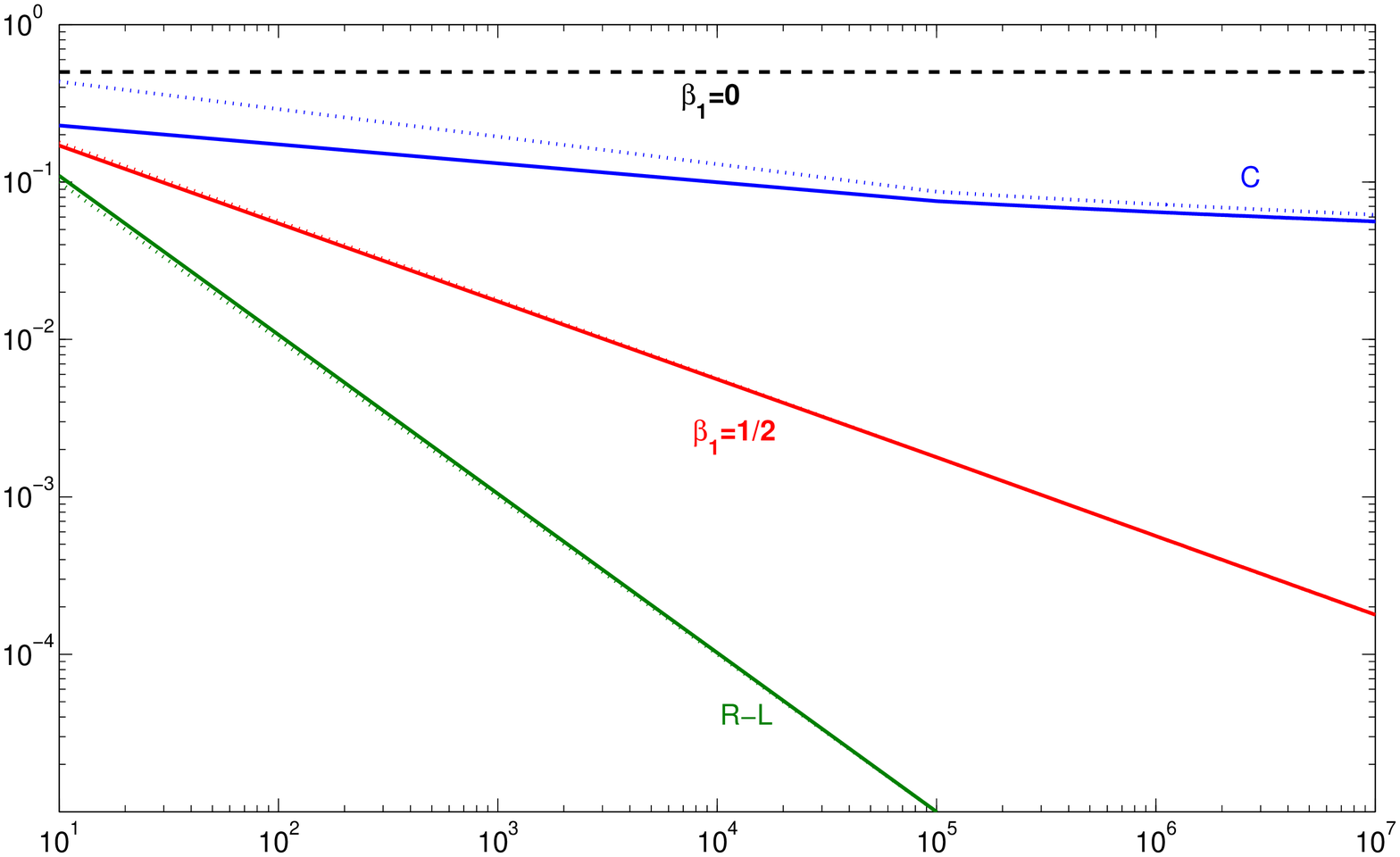}
 \vskip -0.5truecm
\caption{Fundamental solutions for
R-L and C uniformly distributed fractional relaxation
in comparison with some of them for single orders.
Top: linear scales; Bottom: logarithmic scales.}
\end{center}
\end{figure}
\vskip -0.5truecm
\noindent
In Fig. 4 we display
the plots of the fundamental solutions for
R-L and C uniformly distributed fractional relaxation,
adopting as previously,
in the top plate, linear scales ($0 \le t \le  10$),
and in the bottom plate, logarithmic scales ($10^1 \le t\le 10^7$).

For comparison in the top plate the   plots
for single orders $\beta_1= 0, 1/2, 1$ are shown.
 We note
that for $1<t<10$ the R-L and C plots are close to that
for $\beta_1=1/2$ from above and from below, respectively.

In the bottom plate (where the plot for $\beta_1 $
is not visible because of  its faster exponential decay)
we have  added in dotted lines the asymptotic
solutions for large times.
We recognize that the C plot is decaying much slower
than any power law whereas the R-L plot is decaying as
$t^{-1}$; this means that for large times these plots
are the border lines  for all the plots corresponding
to  single order relaxation with  $\beta _1 \in (0,1)$.   
\section{Conclusions}
We have investigated the relaxation equation with (discretely or
continuously) distributed order of fractional derivatives both
in the Riemann-Liouville and in the Caputo sense.
Such equations can be seen as  simple models of more general distributed order
fractional evolution in a Banach space where the relaxation parameter $\lambda$
is replaced by an operator $A$ acting in this space.
A relevant example is time-fractional diffusion where in the linear
case the individual modes exhibit fractional relaxation.
Our interest is focused on structural properties of the solutions,
in particular on asymptotic behaviour at small and large times.
In both  approaches we find that the smallest order of 
occurring fractional differentiation determines
 the behavior near infinity,
but the largest order the behaviour near zero, in analogy to the special form
of time-fractional diffusion explicitly governed by the distributed order derivative
as in \cite{ChechkinGorenfloSokolov PRE02,ChechkinGorenfloSokolovGonchar FCAA03}
and in \cite{Langlands 06,Mainardi-et-al MME-SPRINGER06,Sokolov APP04}.
We see that the two parameters $\beta_1$ and
$\beta_2$ play opposite roles in our two cases (R-L) and (C).
The topic deserves further study in several directions, \eg in
in terms of integral  transforms and special functions like
those of Mittag-Leffler type.

\vskip -1.5truecm
 \section*{Acknowledgements}
  This work has been  carried out in the framework of a  research
  project for {\it Fractional Calculus Modelling} 
 [URL: {\tt www.fracalmo.org}].
 M. Stojanovi\'c appreciates support by German Research Foundation (DFG).

\section*{Appendix: Essentials of fractional calculus}

For a sufficiently well-behaved function $f(t)$
($ t\in \RR^+$) we may define the fractional derivative
of order $\mu  $ ($m-1 <\mu \le m\,,$ $\, m\in \NN$),
 see \eg  \cite{GorMai CISM97,Podlubny 99},
in two different senses,  that we refer here as to
{\it Riemann-Liouville} (R-L) derivative
and {\it Caputo} (C) derivative, respectively.
Both derivatives are related to the so-called Riemann-Liouville
fractional integral of order $\alpha >0$  defined as
$$ _t J^\alpha  \, f(t) :=    \rec{\Gamma(\alpha )}\,
\int_0^t\!  (t-\tau)^{\alpha-1} \, f(\tau )\, d\tau\,, \q \alpha >0\,.
\eqno(A.1)  $$
We recall the convention $\,_tJ^0 = I$ (Identity operator)
and the semigroup property
$$ _tJ^\alpha \, _tJ^\beta = \,
   _tJ^\beta  \, _tJ^\alpha = \, _tJ^{\alpha +\beta} \,, \q
 \alpha ,\beta \ge 0\,. \eqno(A.2)$$
Furthermore
$$ _t J^{\alpha  }\, t^{\gamma}=
   {\Gamma(\gamma +1)\over\Gamma(\gamma +1+ \alpha  )}\,
     t^{\gamma+\alpha  }\,,
 \q  \alpha \ge 0\,,
  \q \gamma >-1\,, \q t>0\,.
\eqno (A.3)
$$
The fractional derivative of order $\mu >0$
in the {\it Riemann-Liouville} sense  is defined as the operator
$\,_tD^\mu$ which is the
left inverse of
the Riemann-Liouville integral of order $\mu $
(in analogy with the ordinary derivative), that is
$$ _tD^\mu \, _tJ^\mu  = I\,, \q \mu >0\,. \eqno(A.4) $$
If $m$ denotes the positive integer
such that $m-1 <\mu  \le m\,,$  we recognize from Eqs. (A.2) and (A.4)
$_tD^\mu \,f(t) :=  \, _tD^m\, _tJ^{m-\mu}  \,f(t)\,, $
hence
$$
 _tD^\mu  \,f(t) = 
 \, \cases{
  {\ds {d^m\over dt^m}}\lt[
  {\ds \rec{\Gamma(m-\mu )}\int_0^t
    \! {f(\tau)\,d\tau  \over (t-\tau )^{\mu  +1-m}} }\rt] ,
 &  $m-1 <\mu  < m,$\cr\cr
   {\ds {d^m \over dt^m} f(t)} \,,
& $ \mu =m.$\cr}
\eqno(A.5)$$
For completion we define $ _tD^0 = I\,. $
On the other hand, the fractional derivative of order $\mu >0$ in the
{\it Caputo} sense  is defined as the operator
$\,_tD_*^\mu$  such that
$    _tD_*^\mu \,f(t) :=  \, _tJ^{m-\mu } \, _tD^m \,f(t)\,,$
hence
$$
    _tD_*^\mu\,f(t) =  
 \, \cases{
    {\ds \rec{\Gamma(m-\mu )}}\,{\ds\int_0^t
 \! {\ds {f^{(m)}(\tau)\, d\tau \over (t-\tau )^{\mu  +1-m}}}} \,,
& $ m-1<\mu  <m,$\cr\cr
   {\ds {d^m \over dt^m} f(t)} \,, & $ \mu =m.$\cr}
\eqno(A.6) $$
Thus, when the order is not integer the two fractional derivatives
 differ in that  the derivative of order $m$
does not generally commute with the fractional integral.
We point out that the  {\it Caputo} fractional derivative
  satisfies the  relevant property
of being zero when applied to a constant, and, in general,
to any power function  of non-negative integer degree less than $m\,,$
if its order $\mu $ is such that $m-1<\mu \le m\,. $
Furthermore we note that   
$$ _t D^{\mu }\, t^{\gamma}=
   {\Gamma(\gamma +1)\over\Gamma(\gamma +1-\mu )}\,
     t^{\gamma-\mu }\,,
 \q \mu  \ge 0\,,
  \q \gamma >-1\,, \q t>0\,. \eqno(A.7)
$$
Gorenflo and Mainardi \cite{GorMai CISM97}
have shown the essential  relationships between the two fractional
derivatives 
(when both of them exist),
$$  _tD_*^\mu  \,f(t)   =  \, \cases{
  {\ds \, _tD^\mu  \,\lt[ f(t) -
  \sum_{k=0}^{m-1} f^{(k)}(0^+)\,{t^k\over k!} \rt]} \,, \cr\cr
 {\ds \, _tD^\mu  \, f(t) -
    \sum_{k=0}^{m-1} {f^{(k)}(0^+) \,
t^{k-\mu }\over \Gamma(k-\mu+1)} }\,,\cr }
\;  m-1 <\mu <m\,. 
\eqno(A.8)$$
In particular, if $m=1$ we have
$$  _tD_*^\mu  \,f(t) = \, \cases{
   {\ds _tD^\mu \,\lt[ f(t) -
   f(0^+) \rt] }\,,\cr
 {\ds _tD^\mu\,f(t) - {f(0^+)\, t^{-\mu} \over\Gamma(1-\mu)}}\,,\cr}
 \q 0<\mu <1\,. \eqno(A.9)$$
The {\it Caputo} fractional derivative,
 practically ignored in the  mathematical treatises,
represents a sort of regularization in the time origin for the
{\it Riemann-Liouville} fractional derivative.
We note that for its existence
all the limiting   values $f^{(k)}(0^+):= {\ds \lim_{t\to 0^+} f(t)}$
are required to be finite for $k=0,1, 2. \dots m-1$.

We observe the different behaviour  of the two fractional derivatives
at the end points of the interval $(m-1,m)\,$
namely when the order is any positive integer:
whereas  $\, _tD^{\mu}$ is, with respect to its order $\mu\,, $
 an operator continuous
 at any positive integer,
$\, _tD_*^{\mu}$   is an operator left-continuous
since
$$ \cases{
{\ds \lim_{\mu \to (m-1)^+}\,_tD_*^{\mu} \,f(t)}\, = \,
  {\ds f^{(m-1)}(t) - f^{(m-1)} (0^+)}\,, \cr
  {\ds \lim_{\mu \to m^-}\,
 _tD_*^\mu  \,f(t)}\, = \,  {\ds f^{(m)}(t)}\,.}
 \eqno(A.10)
$$
We also note for $m-1 < \mu \le m\,,$
$$   _t D^\mu  \, f(t) \,=\, _tD^\mu   \, g(t)
   \,  \Longleftrightarrow  \,
  f(t) = g(t) + \sum_{j=1}^m c_j\, t^{\mu -j} \,,  \eqno(A.11)
    $$
$$    _tD_* ^\mu  \, f(t) \,=\,  _tD_*^\mu   \, g(t)
   \,  \Longleftrightarrow  \,
  f(t) = g(t) +  \sum_{j=1}^m c_j\, t^{m-j} \,. \eqno(A.12)
     $$
In these formulae the coefficients $c_j$ are arbitrary constants.
Last but not least, we point out the major utility
of the Caputo fractional derivative
in treating initial-value problems for physical and engineering
applications where initial conditions are usually expressed in terms of
integer-order derivatives. This can be easily seen
using the Laplace transformation, according to which
$$ \L \lt\{ _tD_*^\mu \,f(t) ;s\rt\} =
      s^\mu \,  \widetilde f(s)
   -\sum_{k=0}^{m-1}    s^{\mu  -1-k}\, f^{(k)}(0^+) \,,
  \q m-1<\mu  \le m \,,\eqno(A.13) $$
where
$ \widetilde f(s) =
\L \lt\{ f(t);s\rt\}
 = {\ds \int_0^{\infty}} \! \e^{\ds \, -st}\, f(t)\, dt\,, \;
s \in \CC$, and  $ f^{(k)}(0^+) := {\ds \lim_{t\to 0^+}}\, f(t)$.
The corresponding rule for the Riemann-Liouville
derivative is more cumbersome:  for $m-1<\mu \le m $ it reads
$$ \L \lt\{ _tD^\mu  \, f(t);s\rt\} =
      s^\mu \,  \widetilde f(s)
   -\sum_{k=0}^{m-1}\,
\lt[_tD^k\, _tJ^{(m-\mu)}\rt]\,f(0^+) \, s^{m -1-k}\,,
\eqno(A.14)$$
where, in analogy with (A.13),  the limit for $t \to 0^+$
is understood to be taken after the operations of fractional integration
and derivation.  As soon as all the limiting   values $f^{(k)}(0^+)$
are finite
and $m-1 <\mu< m$, 
the formula (A.14)  simplifies into
$$ \L \lt\{ _tD^\mu  \, f(t);s\rt\} =
      s^\mu \,  \widetilde f(s) \,.\eqno(A.15)$$
In the special case   $f^{(k)}(0^+)=0$  for $k=0,1,  m-1$,
we recover the identity between the two fractional derivatives,
consistently with Eq. (A.8).

For more details
on the theory and applications of fractional calculus
we recommend
to consult in addition to the
well-known books by Samko, Kilbas \& Marichev \cite{SKM 93},
by Miller \& Ross \cite{Miller-Ross 93},  
by Podlubny \cite{Podlubny 99},
 those appeared in the last few years,  
by  Kilbas, Srivastava \& Trujillo \cite{Kilbas-et-al BOOK06},
by West, Bologna \& Grigolini \cite{West BOOK03},
and by Zaslavsky \cite{Zaslavsky BOOK05}.

\end{document}